\documentclass[12pt]{article}
\usepackage{graphicx}
\usepackage{amsmath}
\usepackage{colordvi}
\usepackage{latexsym,graphicx,color,times}
\usepackage{geometry}
\newcommand{\be}{\begin{equation}}
\newcommand{\ee}{\end{equation}}

\def\Alf{Alfv\'en }

\def\gfac{4}
\def\gfac{3.4}
\def\Red{\Cyan}
\def\Cyan{\Black}
\def\kapl{\chi_{\parallel}}
\def\taup{\tau_{\parallel}}
\def\req#1{(\ref{eq:#1})}
\def\rfig#1{Fig.~\ref{fig:#1}}
\def\rsec#1{Sec.~\ref{sec:#1}}
\def\macname{hank_strauss}

\def\figdir3d{/Users/{\macname}/Documents//progs/m3dc1/d3d}

\begin{document}
\begin{center}
{\bf\large Resistive wall tearing mode disruptions in DIII-D  and ITER tokamaks} \\
{H. R. Strauss$^{1 ,} $ \footnote{Author to whom correspondence should be addressed: hank@hrsfusion.com}, {B.C. Lyons}$^2$, M. Knolker$^2$} \\
{$^1$ HRS Fusion, West Orange, NJ 07052 } \\
{$^2$ General Atomics, San Diego, CA 92121}
\end{center}

\abstract{
Disruptions are a serious problem in tokamaks, 
in which thermal and magnetic energy confinement  is lost.
This paper 
 uses data from the DIII-D experiment, theory, and simulations to demonstrate that
resistive wall tearing modes (RWTM)  produce the thermal quench (TQ) in a
typical locked mode shot.
\Cyan{Analysis of the linear RWTM dispersion relation shows 
the parameter dependence
of the growth rate, particularly on the resistive wall time. 
Linear simulations of the locked mode equilibrium
show that it is unstable with a resistive wall, and stable with an
ideally conducting wall. 
Nonlinear simulations 
demonstrate that the RWTM grows to sufficient amplitude to cause a complete 
thermal quench. The RWTM growth time is proportional to the
thermal quench time.
The nonlinearly saturated RWTM magnetic perturbation
amplitude agrees with experimental measurements. 
The onset condition
is that the
$q = 2$ rational surface is sufficiently close to  the resistive wall.
Collectively, this identifies 
the RWTM as the cause of the TQ.} 
In ITER, RWTMs will   
produce long TQ times compared to present-day experiments. 
ITER disruptions may be significantly more benign than previously  
predicted.
}

\section{Introduction}
Disruptions are a serious problem in tokamaks, 
in which thermal and magnetic energy confinement  is lost.
\Cyan{This is thought to be a severe problem in large tokamaks such as
ITER.}
It was not known what instability causes the thermal quench (TQ)  in disruptions, and how
to avoid it.  
Recent work identified the thermal quench  in JET locked mode disruptions  with
a resistive wall tearing mode (RWTM) \cite{jet21}. The RWTM was also predicted
in ITER disruptions \cite{iter21}.
This paper 
 uses data from the DIII-D experiment, theory, and simulations to demonstrate that
resistive wall tearing modes cause the thermal quench in a typical locked mode shot.

\begin{figure}[h]
\begin{center}
\includegraphics[width=5.5cm,width=10cm]{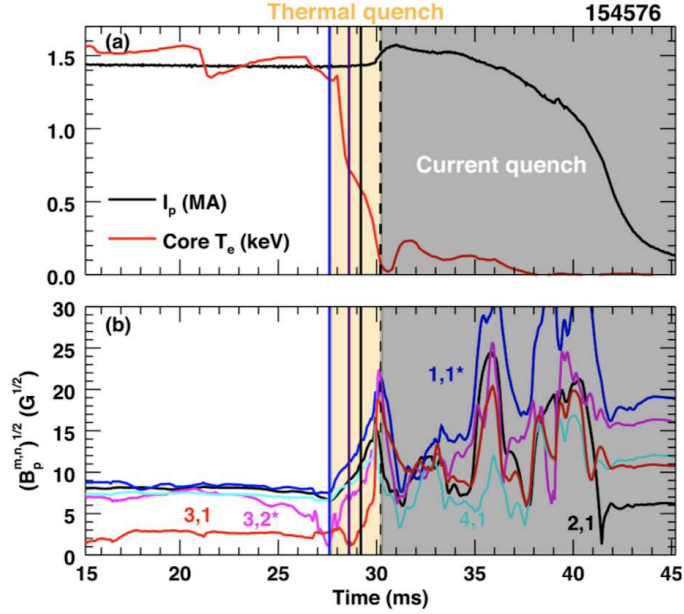}
\end{center}
\caption{\it
DIII-D shot 154576 \cite{sweeney}, where $T_e$ in the upper frame  is the 
temperature on a core magnetic 
surface, and $I_p$ is the toroidal current. The lower frame shows magnetic probe signals,
dominated by $n=1$  toroidal harmonics.
The  TQ time is
$\tau_{TQ} = .5 \tau_{wall} = 2.5 ms,$ 
and $\tau_{wall} = 5 ms$.
Reprinted from \cite{sweeney} with IAEA permission.}
\label{fig:d3data}
\end{figure}

\Cyan{
There can be many sequences of events leading to a disruption \cite{devries12},
which generally culminate in a locked mode.}
In locked mode shots, the plasma is at first  toroidally rotating and
 relatively quiescent. 
The rotation slows, and the plasma becomes unstable to tearing modes. 
\Cyan{
Locked modes are the main precursor of JET
disruptions, but they are not the instability causing the
thermal quench.  Rather,
the locked mode indicates an ``unhealthy" plasma which may disrupt
\cite{gerasimov2020}.
It was conjectured that at a critical amplitude, tearing modes would
overlap and cause a disruption \cite{devries16}, 
 although in many cases, such as  in \rfig{d3data}, 
 the mode amplitude does not increase before the TQ occurs.
 Prior to and during the locked mode, the plasma can develop 
low   current and temperature 
outside the $q = 2$ surface. This can be caused by tearing mode island overlap 
\cite{schuller} and impurity radiation \cite{pucella}.
After the  TQ, a current quench (CQ) occurs, as in \rfig{d3data},  
caused by high plasma resistivity.}



In the following we  discuss
experimental data of a locked mode disruption in DIII-D,
linear theory of resistive wall tearing modes, 
linear and nonlinear simulations, and the onset conditions of the RWTM.
The experimental data shows that the TQ occurs in about half a resistive
wall time.
Linear theory presented in \rsec{linear}  shows the connection of tearing modes 
and resistive wall tearing modes. The scaling with resistive wall time is found.
Linear simulations show that the mode is stable for an ideal wall, and
unstable with a resistive wall. \Cyan{ 
}
Nonlinear simulations in \rsec{nonlinear} show that the
mode grows to large amplitude, sufficient to cause a complete thermal quench.
\Cyan{
The thermal quench time is proportional to the reciprocal of the mode growth rate.
 The mechanism of the thermal quench  is shown to be parallel thermal transport.
The peak amplitude of the magnetic perturbations agrees with experimental measurement.}
The mode onset occurs when  the radius of the $q =2$ resonant surface 
is sufficiently close to  the plasma edge as shown in \rsec{onset},
\Cyan{consistent with a database of DIII-D disruptions
\cite{sweeney2017}.}

\Cyan{A summary and the implications for ITER are given in \rsec{summary}. 
It is not known whether ITER will rotate and lock, but
it can have RWTMs \cite{iter21}.
The long timescale of the TQ implies that the
mitigation requirements of ITER disruptions \cite{lehnen} could be greatly relaxed.}

\rfig{d3data} shows data  from 
DIII-D shot 154576
  \cite{sweeney}. A locked mode persists until the thermal quench. 
The TQ occurs in the time range $27.5 - 30 ms,$ marked with vertical
lines. The upper frame shows the temperature $T_e$ on a core flux surface.  
Also shown is the toroidal current $I_p$, which spikes after the TQ, and then  quenches on a slower
timescale.
The lower frame shows magnetic probe signals, dominated by  the   $n = 1$ toroidal
harmonic.
\Red{The harmonics of the magnetic field are mapped to their
respective rational surface \cite{sweeney} .}
The TQ time is 
$\tau_{TQ} \approx 2.5 ms \approx .5 \tau_{wall}$ where the resistive wall penetration time
is $\tau_{wall} = 5 ms$.  The TQ time is {proportional to} 
the experimental growth time
of the $n = 1$ magnetic perturbations. 
Before the TQ occurs, there is a locked mode,  which consists of low 
amplitude precursors, identified
as tearing modes (TM) \cite{sweeney}.

This is similar to
JET shot 81540 \cite{jet21}, with  TQ time
$\tau_{TQ} = .3 \tau_{wall} = 1.5 ms .$
and $\tau_{wall} = 5 ms$.
The growth time of the mode is \Red{proportional to} the  TQ time, indicating the mode growth causes
the TQ.  
These results 
suggest that a resistive wall mode  (RWM) or resistive wall tearing mode (RWTM) causes the TQ.

\section{Linear stability} \label{sec:linear}
\Cyan{The RWTM
  dispersion relation generalizes the standard
tearing mode by taking into account diffusion of the magnetic perturbation
across the resistive wall.  This can be expressed for a thin wall as
$ \gamma \psi = (\eta_w/\delta_w) ( \psi'_{vac} - \psi' ) $\label{eq:thin0} 
where $\psi$ is the magnetic potential at the wall, 
$\psi'$ is its radial derivative on the plasma side of the wall, 
$\gamma$ is the growth rate, 
$\eta_w,\delta_w$ are the wall resistivity and thickness, 
and $\psi'_{vac} = - (m/r_w) \psi$ is the radial derivative of $\psi$ on the vacuum
side of the wall, 
where $r_w$ is the wall radius and 
$m$ is the polodal mode number. This can be expressed 
\be \left(1 + \frac{\gamma \tau_{wall}}{m} \right)\psi  = - \frac{r_w}{m} \psi',\label{eq:thin}\ee
giving a logarithmic derivative boundary condition  for $\psi$ at  a resistive wall,
instead of $\psi = 0$ at an ideal wall.
The resistive wall penetration time is 
$\tau_{wall} = r_w \delta_w / \eta_w.$ 
The RWTM dispersion relation is derived like that of  a tearing mode, using the boundary condition
\req{thin}.}

The dispersion relation \cite{jet21,finn95} is 
\be \hat{\gamma}^{5/4} {S}^{3/4} = \Delta_{i} + \frac{\Delta_{x}}
{\hat{\gamma}{S}_w  +  1 } \label{eq:disp}  \ee
where $\hat{\gamma}  = \gamma \tau_A,$ $S$ is the Lundquist number,
$\tau_A = R / v_A$ is the \Alf time, $R$ is the major radius, $v_A$ is the
\Alf speed,
$S_w = S_{wall}(1 - x_s^{2m})/(2m),$ $S_{wall} = \tau_{wall}/\tau_A,$
ideal wall stability parameter $\Delta_{i} = r_s \Delta_{ideal}'/m,$
external stability parameter $\Delta_{x} = 2 x_s^{2m} / (1 - x_s^{2m})$,
$x_s = (r_s/r_w),$
with rational surface radius $r_s$.

\begin{figure}[h]
\begin{center}
\includegraphics[width=7.0cm]{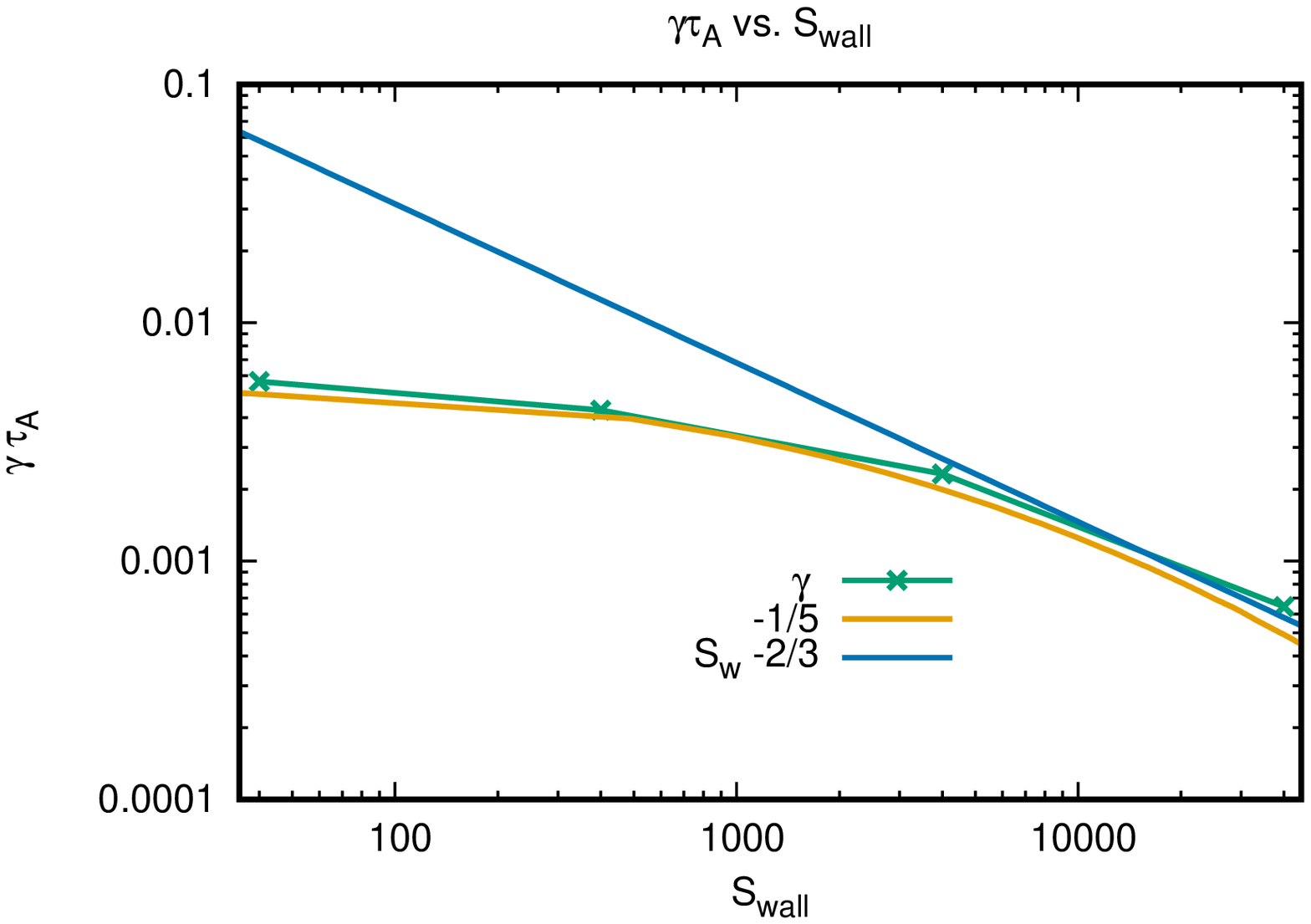}(a)
\includegraphics[width=3.0cm]{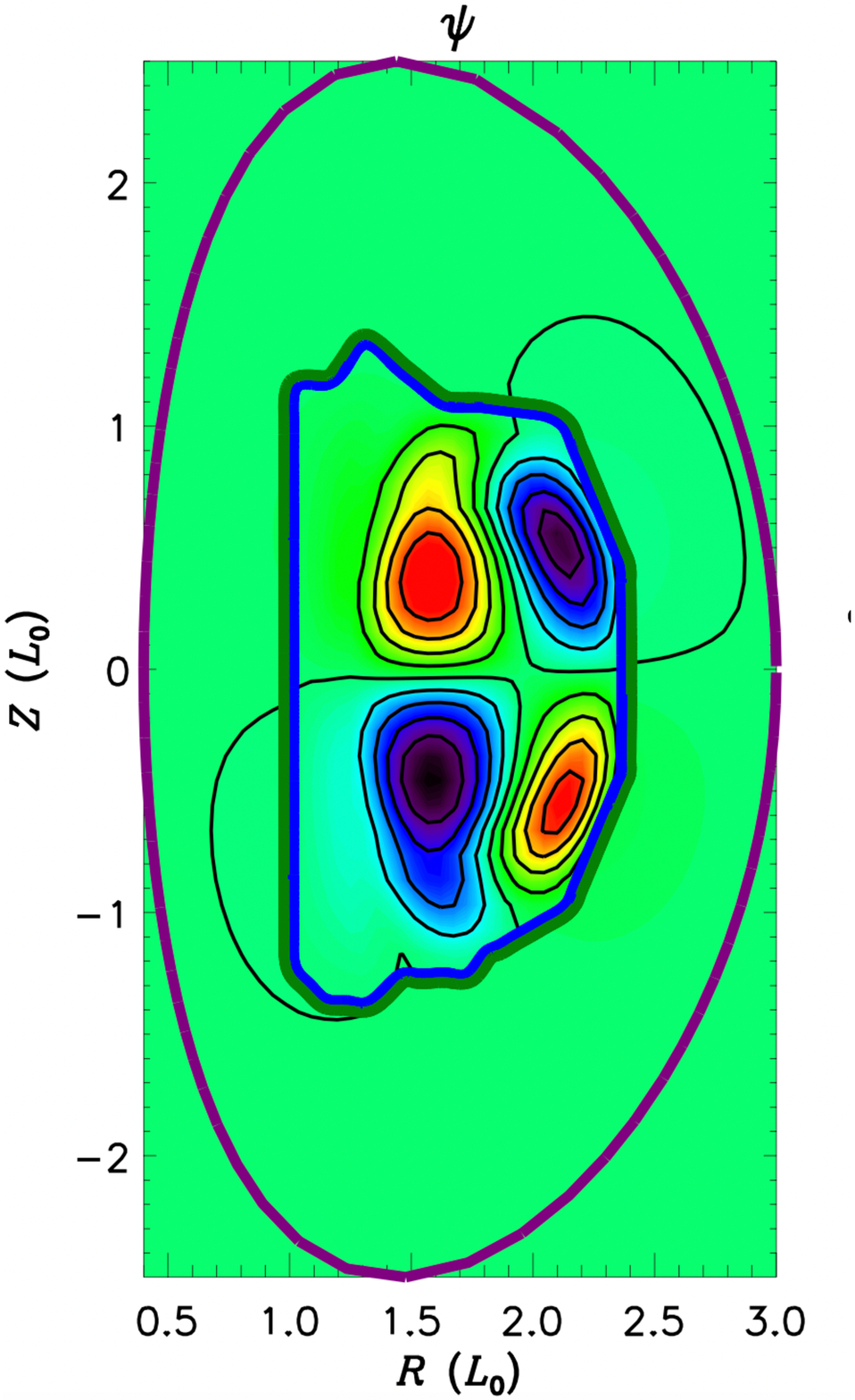}(b)
\end{center}
\caption{\it
(a) $\gamma \tau_A $ in DIIID shot 154576 as a function of  $S_{wall}$
from M3D-C1 linear simulations. 
The fit is  to  a RWTM with \Red{ $\Delta_x = 1,$ $\Delta_i = -1/5,$
from the linear dispersion relation,
which asymptotes to $S_{wall}^{-2/3}.$} 
(b) perturbed $\psi$ in (a). The mode is $(2,1)$,
and penetrates the resistive wall. }
\label{fig:m3dc1ln}
\end{figure}

\Cyan{Ideal wall tearing modes are obtained from \req{disp}
in the limit $S_w \rightarrow \infty,$ 
while \Cyan {modes with a highly resistive wall}  are obtained 
in the limit $S_{w} \rightarrow 0,$
\be \hat{\gamma} = [(\Delta_i + \Delta_x )]^{4/5} S^{-3/5}. \label{eq:tm}  \ee}
Resistive wall tearing modes have $\Delta_{i} \le 0,$
and require finite $S_w.$
The RWTM growth rate scalings can be approximated  from \req{disp}. 
If $\Delta_{i} = 0,$ then 
assuming $\hat{\gamma}{S}_w  \gg  1$
\cite{jet21},
\be \hat{\gamma} = \Delta_x^{4/9} S^{-1/3} S_w^{-4/9}. \label{eq:jet} \ee
If $\Delta_{i} < 0$ and $ \Delta_{x} + \Delta_{i} > 0$, then neglecting the left side of \req{disp} gives
a kind of RWM with rational surface in the plasma, 
\be \hat{\gamma} = - \left(1 + \frac{\Delta_x}{\Delta_i} \right) S_w^{-1}. \label{eq:rwm} \ee
If $\Delta_i + \Delta_x < 0$ there are no unstable solutions of \req{disp}.
Intermediate  asymptotic scalings of $\hat{\gamma}$  are possible depending on the ratio
$\Delta_{i} / \Delta_{x},$ as in \rfig{m3dc1ln}(a). 

\Cyan{The dispersion relation can  include a generalized Ohm's law in the
tearing layer, including diamagnetic drifts. These effects are expected to be small,
particularly in the limit \req{rwm} which does not contain the resistivity.} 
Toroidal rotation is known to stabilize
these modes
\cite{gimblett,bondeson,betti}.
The required rotation frequency is comparable to the
 TM linear growth rate.
After mode locking, the residual rotation is not enough to stabilize the mode.

We now turn to numerical solutions of the equilibrium reconstruction of DIII-D shot 154576.
Linear stability 
was studied using the M3D-C1 \cite{m3dc1}  code,
with a resistive wall \cite{ferraro}. 
The reconstruction had on axis safety factor $q_0 > 1$ to 
prevent the $(1,1)$ mode from dominating the simulations.
\Cyan{It represents the equilibrium just before the TQ. 
The equilibrium is axisymmetric and does not include magnetic
islands.}
\rfig{m3dc1ln}(a) shows the growth rate, as a function of $S_{wall}.$ 
The curve  labelled \Red{$-1/5$  
has $\Delta_x = 1,$ $\Delta_i = -1/5$, which asymptotes to  $\gamma \propto S_{wall}^{-2/3}$.}
The $S_{wall}=0$ limit is  a tearing mode with \Cyan{highly resistive wall}.
The numerical solution is intermediate between the scalings  \req{jet} and  \req{rwm}.

\rfig{m3dc1ln}(b) shows the perturbed magnetic flux $\psi,$ showing a $(2,1)$ structure.
When the wall is ideally 
conducting, the mode is stable.
This shows that the  mode is not an ideal wall TM. It must necessarily have
$\Delta_i \le 0.$

\section{Nonlinear simulations} \label{sec:nonlinear}
The linear simulations establish that the equilibrium reconstruction is unstable to
a RWTM, and stable to an ideal wall TM. Nonlinear simulations show that the mode grows to 
large amplitude, sufficient to cause a thermal quench.
\Cyan{The simulations were performed  with M3D \cite{m3d} with a thin 
resistive wall \cite{pletzer}.
The simulations used the same parameters as \cite{jet21,iter21}, in particular  $S= 10^6$,
parallel thermal conductivity $\chi_\parallel = 10 R^2 / \tau_A,$ perpendicular 
thermal conduction and viscosity $\chi_\perp = \mu = 10^{-4} a^2 / \tau_A,$
where $a$ is the minor radius. 
The simulations used $16$ poloidal planes, adequate to
resolve low toroidal mode numbers. The simulations and the experimental data were
dominated by $n = 1$ modes.
}

\begin{figure}[h]
\begin{center}
\includegraphics[height=6.5cm]{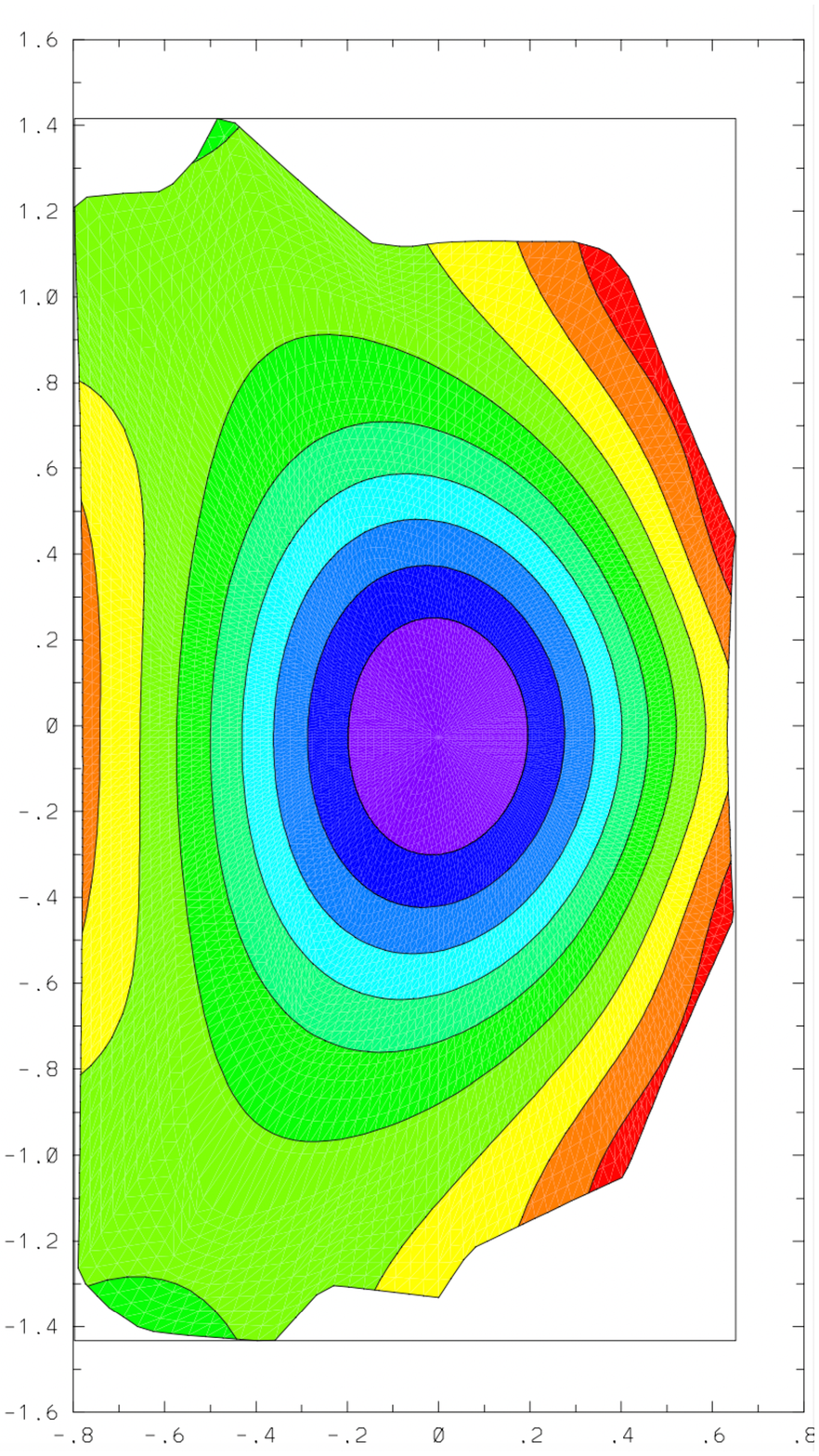}(a)
\includegraphics[height=6.5cm]{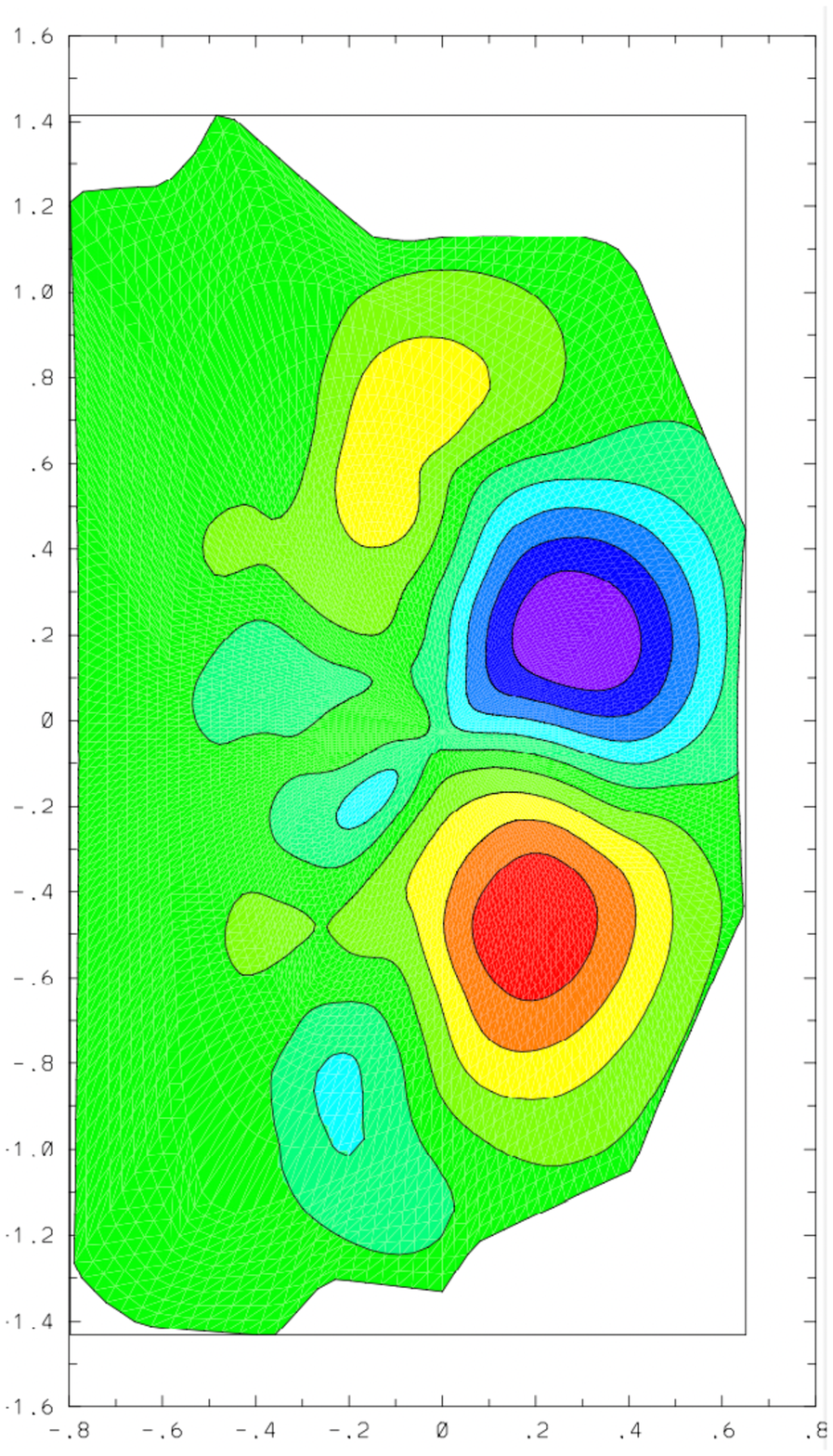}(b)
\includegraphics[height=6.5cm]{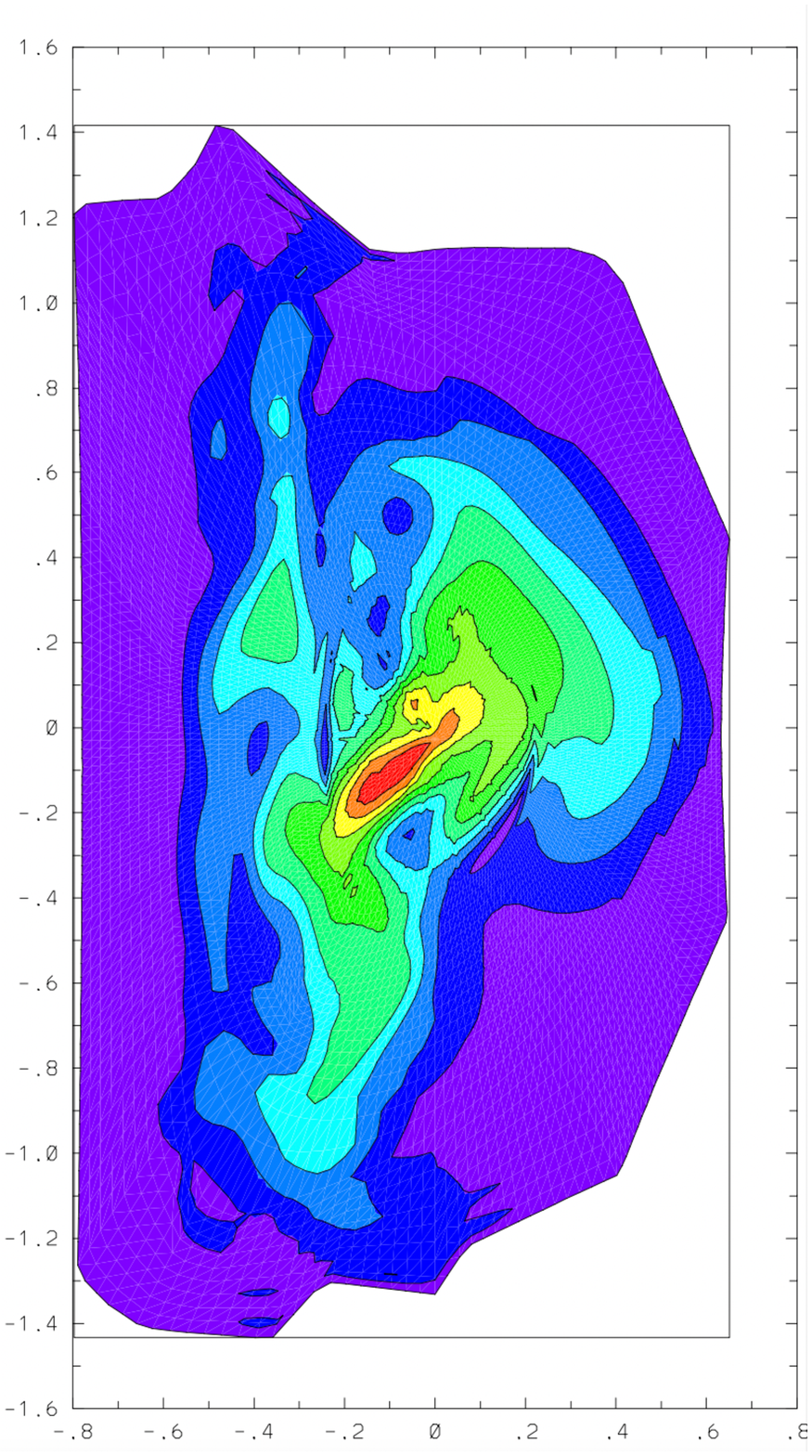}(c)
\end{center}
\caption{\it
(a) initial $\psi$ of  DIII-D 154576.
(b) perturbed $\psi$ at $t = 5690 \tau_A,$ $S_{wall} = 10^4.$
(c) $p$ at $t = 5690 \tau_A.$
when $P $ is about 20\% of its initial value.}
\label{fig:nonlin1}
\end{figure}

\rfig{nonlin1} shows a simulation with M3D \cite{m3d} with a resistive wall \cite{pletzer},
  of the same equilibrium reconstruction of DIII-D 154576.
The simulation had $S_{wall} = 10^4.$ 
Experimentally, $S_{wall} = 1.2 \times 10^4.$
The initial magnetic flux $\psi$ is shown in \rfig{nonlin1}(a), and the
perturbed $\psi$ is in \rfig{nonlin1}(b), at a time late in the simulation,
when the TQ is almost complete. The nonlinear perturbed $\psi$ is predominantly
$(2,1)$, similar to  the linear
structure of \rfig{m3dc1ln}(b).
The pressure, shown at the same time in \rfig{nonlin1}(c) 
has a  large perturbation that causes the TQ. 
The pressure is shown when the total volume integral of the  pressure 
 $P$ is about 20\% of its initial value.

\begin{figure}[h]
\begin{center}
\includegraphics[height=5.0cm]{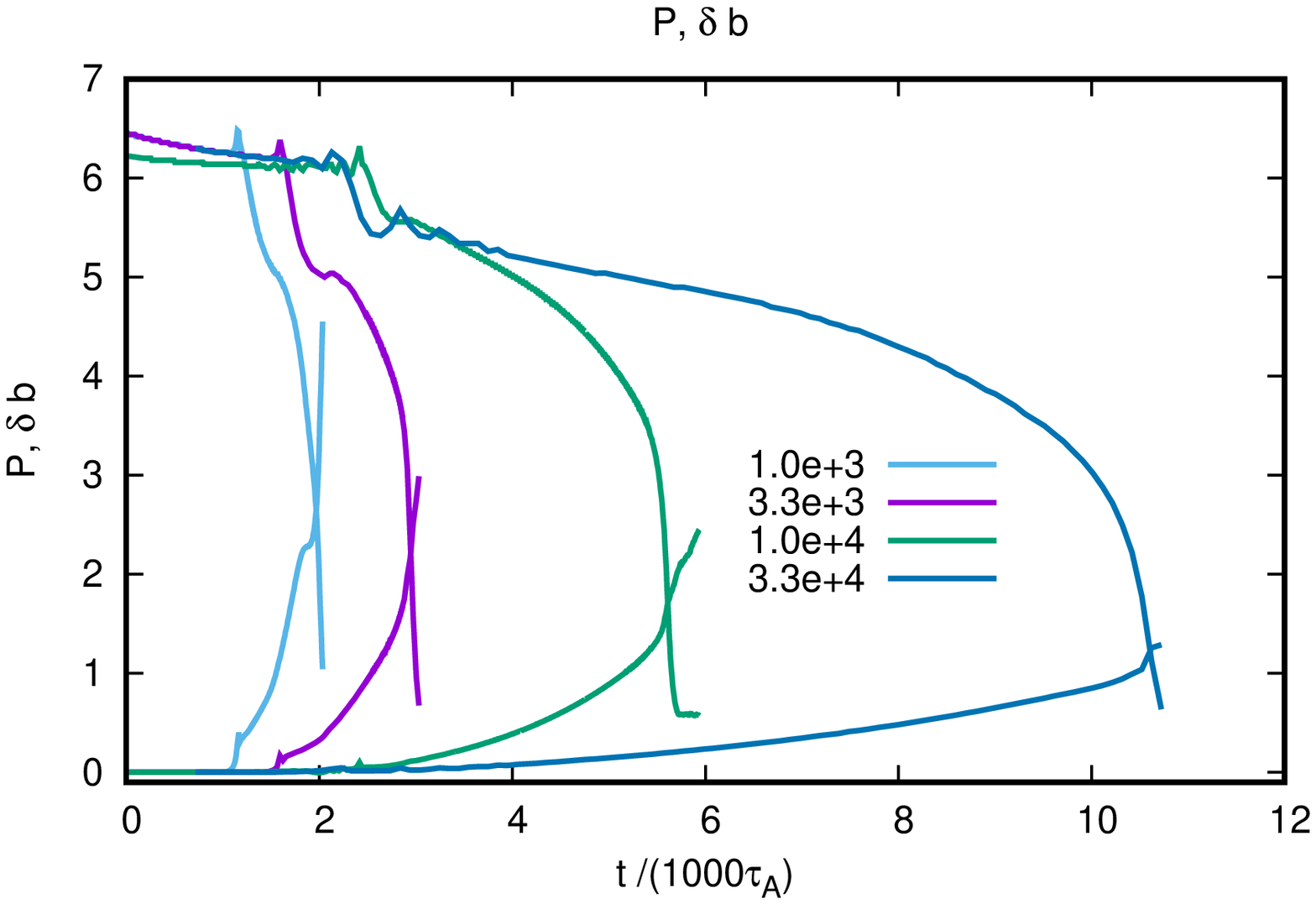}(a)
\includegraphics[height=5.0cm]{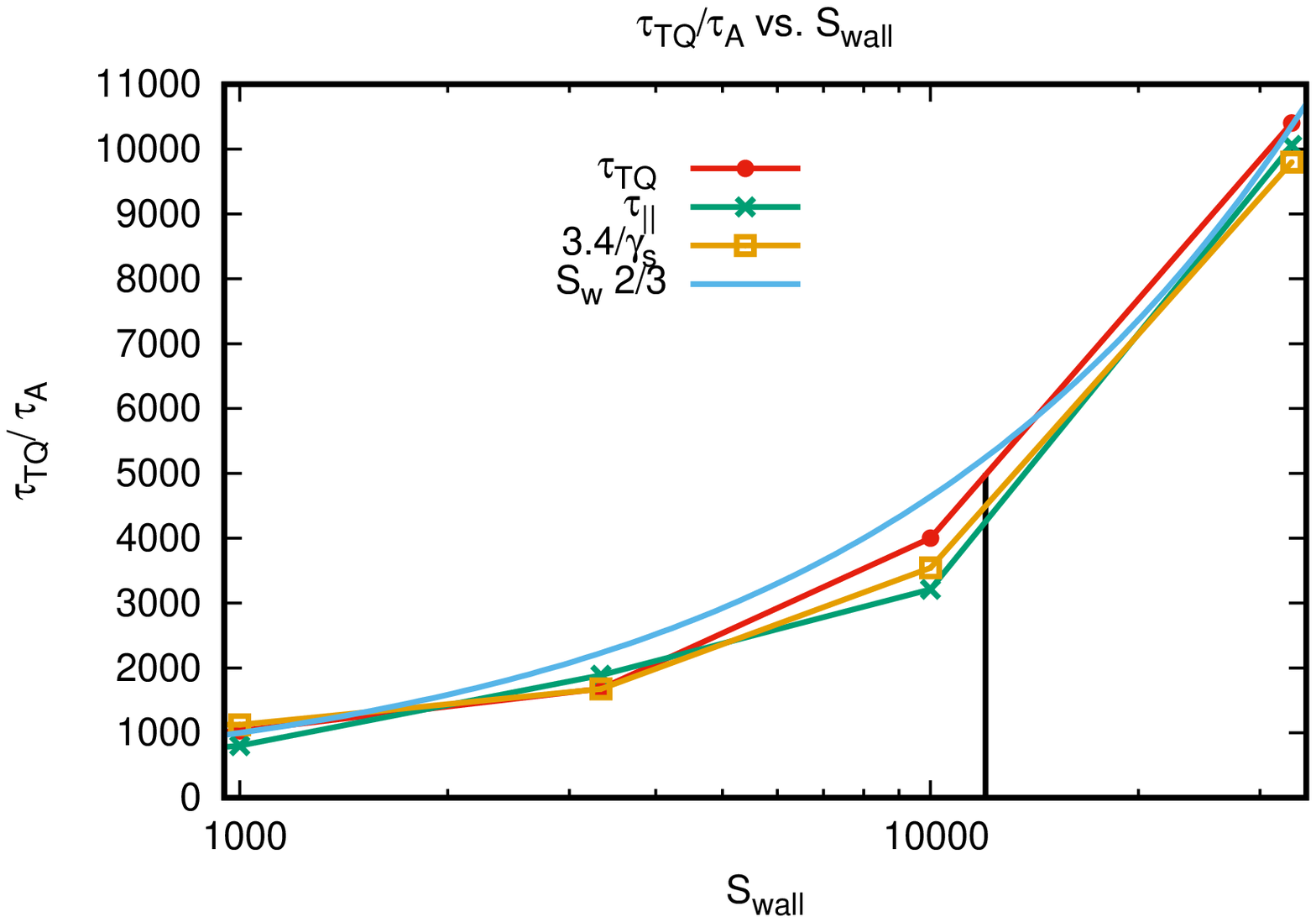}(b)
\end{center}
\caption{\it
(a) time histories of $P$ and $b_n$ in M3D simulations of  DIII-D 154576,
where $P$ is total pressure, $b_n$ is perturbed normal $\delta B/B$ at the wall.
\Cyan{The  curves are labeled by the value of $S_{wall}.$}
(b) TQ time $\tau_{TQ}$ measured from the time histories of $P$. 
\Red{ Also shown is $\taup,$ the parallel transport time \req{taup}. The fits are to
$ \gfac/\gamma_s,$ where $\gamma_s$ is measured
from the time histories of $b_n$  in \rfig{nonlin2}(a), and to $S_{wall}^{2/3}.$} 
 The vertical line is the
experimental value of $S_{wall}.$ At the vertical line 
\Red{$\tau_{TQ} / \tau_A \approx 0.4 S_{wall},$ consistent with experimental data.}
}
\label{fig:nonlin2}
\end{figure}

\begin{figure}[h]
\vspace{.5cm}
\begin{center}
 \includegraphics[height=5.0cm]{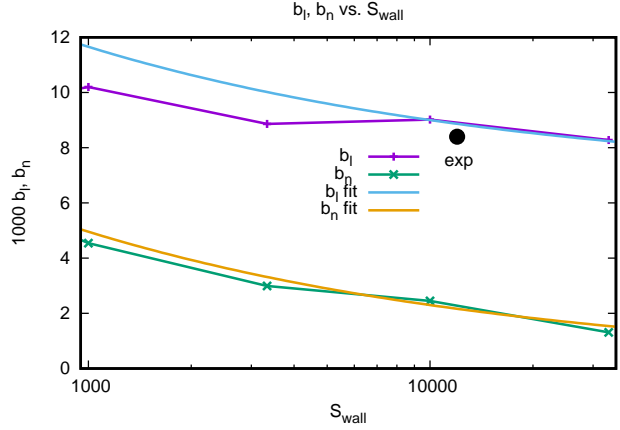}
\end{center}
 \caption{\it \Cyan{
Peak values  of  $b_l$ and $b_n$ in units of $10^{-3},$ as a function of 
$S_{wall}$ in the simulations of
\rfig{nonlin2}(b). 
The  fits are to  $b_n =  0.05 S_{wall}^{-1/3},$ 
$b_l = b_n + 6.7\times 10^{-3}.$} The experimental value is indicated.}

\label{fig:bw}
\end{figure}

\rfig{nonlin2} shows several M3D simulations with different values
of $S_{wall}$. \rfig{nonlin2}(a) shows time histories of $P$ and 
$b_n,$ 
where 
$b_n$ is the perturbed normal $\delta B/B$ at the wall.
\Cyan{
The curves are labelled by the value of $S_{wall}.$ 
All mode numbers $n > 0$ are included in $b_n.$}
\rfig{nonlin2}(b) shows the TQ time $\tau_{TQ}$ measured from the time histories of $P$.
\Red{Also shown is $\taup,$ the parallel transport time,
given by \cite{jet21,iter21} 
\be \taup = \frac{a^2}{\kapl b_n^2} \label{eq:taup} \ee
using the maximum value of $b_n$ for each $S_{wall}$ in \rfig{nonlin2}(a).}
\Red{The fits are to
$\gfac/\gamma_s,$ where the growth rate  $\gamma_s$ is measured
from the time histories of $b_n$  in \rfig{nonlin2}(a), and 
to $S_{wall}^{2/3}.$ This agrees with \rfig{m3dc1ln}(a). The relation
$\taup \propto S_{wall}^{2/3}$
gives the scaling
$b_n \propto S_{wall}^{-1/3},$ 
which also agrees with \rfig{bw}.}
The vertical line is the
experimental value of $S_{wall}.$ \Red{At the vertical line $\tau_{TQ} / \tau_A \approx 0.4 S_{wall}
,$ or $2 ms,$  consistent with  experimental data.}
The mode growth occurs on the same time scale as the TQ, as in the experiment.
The small drop in $P$ in \rfig{nonlin2}(a) at $t \approx 3000\tau_A$ is due to internal modes.
This resembles the minor precursor disruptions observed in
JET \cite{jet21} and DIII-D \cite{sweeney}, which can be seen in \rfig{d3data}. 

The reason the mode grows to large  amplitude may  be  the external stability parameter
$\Delta_x.$ TMs have 
internal stability parameter which depends on the current profile. 
Growth of an island flattens the current gradient and 
stabilizes the TM  at a moderate amplitude.
The external stability parameter $\Delta_{x}$
depends  only on $x_s,$ independent of island size.
It is not saturated by local flattening of the current profile.
It saturates by driving the $q = 2$ surface to the origin, $r_s = 0.$
This is evident from the $m = 2$ structure in $p$ near the axis in \rfig{nonlin1}(c).


\Cyan{The experimental value of $\delta B / B$ is in 
agreement with the simulations.
At its maximum value, the $(2,1)$ magnetic perturbation in \rfig{d3data} is 
$\delta B \approx 480 G,$ 
or $\delta B / B = 2.8 \times 10^{-2},$
taking $B = 1.7 T.$
The value of $\delta B$, measured by probes,  was 
estimated  \cite{sweeney}
at the $(2,1)$ rational surface $r_s$, as noted in the discussion of \rfig{d3data} above.  
To obtain its value
$b_w$ at the wall, $\delta B / B$ must be multiplied by
$x_s^3 = 0.3,$ where $x_s = .67,$    yielding $b_w = 8.4 \times 10^{-3}.$
 To compare with the simulation, 
\rfig{bw} shows $b_l$, the peak transverse perturbed magnetic
field at the wall as a function of $S_{wall},$  and peak $b_n$, also shown in \rfig{nonlin2}(a),  
in units of $10^{-3}.$ The $b_n$ signal would be measured by saddle coils in the experiment,
while $b_l$ would be measured by probes, like $b_w.$
The peak value of $b_l$ can be  fit using \req{thin}, with
$(b_n,b_l) \propto (m\psi /r_w, \psi')$, 
noting that 
$ \gamma \tau_{wall} b_n = const.$ The fit is
$b_n \approx 0.05 S_{wall}^{-1/3},$ $b_l = b_n  + 6.7 \times 10^{-3}.$  
The maximum value $b_l = 8.8 \times 10^{-3}$ at the experimental value 
$S_{wall} = 1.2\times 10^4$, is in agreement with $b_w.$
The experimental value is indicated in \rfig{bw}.}

\section{Onset condition} \label{sec:onset}
Having shown that the equilibrium reconstruction is unstable to  a RWTM that grows large
enough to produce a TQ,
we consider the onset 
 condition for the RWTM, 
 $\Delta_{i} = 0.$  
Both $\Delta_{x}$ and $\Delta_{i}$ depend on $x_s.$ 
In a step current model \cite{finn95,frs73} with 
a constant current density contained within radius $r_0$,
 with $r_s / r_0 = (2 / q_0)^{1/2},$ 
and $(m,n) =(2,1)$,
\be \Delta_{i} = -2 \frac{q_0 - 1 - (q_0^2/4) x_s^4}
{[q_0 - 1 - q_0^2/4](1 - x_s^4)} \label{eq:delta1} \ee
For $\Delta_{i} > 0$ this requires a positive numerator and negative denominator.
If this is not satisfied, the RWTM is unstable,
\be x_s^4 > (4/q_0^2)(q_0 - 1). \label{eq:xs4} \ee
For example, if $q_0 = 1.05,$ then $\Delta_{i} = 0$ for $x_s = 0.65.$ For larger $x_s>  0.65$,
the RWTM is unstable.
Other current profiles are similar \cite{frs73}, in that
larger $x_s$ causes $\Delta_{i} \le 0 $.
There is additional experimental support for this.
Whether a disruption occurs or not depends on the normalized  $q = 2$
radius
$\rho_{q2}$ in a database of DIII-D 
locked modes  
\cite{sweeney2017}
Here $\rho_{q2} \approx  r_s / a,$ 
and $r_{w} / a \approx 1.2.$
The disruption onset boundary in the database is $\rho_{q2} > 0.75,$ or
$x_s > 0.625.$
In the simulations, $\rho_{q2} = 0.8,$
so that
$ x_s = 0.67. $ 
The onset condition is that the $q = 2$ surface is close enough to the plasma edge.
The database also shows that $\rho_{q2}$ increases in time
from the beginning to the end of the locked  mode,
as the disruption is
approached. This connects the onset condition  to profile evolution.

\section{Summary and implications for ITER} \label{sec:summary}


To summarize, 
theory and simulations were presented of resistive wall tearing modes,
in an equilibrium reconstruction of DIII-D shot 154576. 
The linear tearing mode  dispersion relation with a resistive wall 
showed the parameter dependence of the modes, especially on $S_{wall}.$
Linear simulations found that
the equilibrium  was stable with an ideally conducting wall,
and unstable with a resistive wall. 
\Cyan{For the  particular example studied here, the growth rate scales as
$\gamma \propto S_{wall}^{-2/3}.$} 
The RWTMs grow to
large amplitude nonlinearly. 
\Cyan{
The thermal quench time is proportional to the RWTM growth time.
The amplitude of
the simulated peak magnetic perturbations is in agreement with the experimental data.}
The onset condition for disruptions is that the 
$q = 2$ rational surface is close enough to the plasma edge,
consistent with a DIII-D disruption database.

These results are very favorable for ITER disruptions.
The ITER resistive wall time, $250 ms,$ is $50$ times
longer than in JET and DIII-D. The TQ time, instead of being $1.5 - 2.5 ms$ in
JET and DIII-D respectively, could be $75  - 125 ms,$
assuming the TQ is produced by a RWTM with {$\tau_{TQ} \approx \tau_{wall}/2$} scaling.
If the TQ is caused by a
RWTM with $S_{wall}^{-4/9}$, and the edge temperature is $500 eV,$
then \cite{iter21}  $\tau_{TQ} = 70 ms.$
The highly conducting ITER wall strongly mitigates RWTMs. It could  greatly
relax the requirements of the disruption mitigation system, disruption
prediction, and mitigation of runaway electrons.

{\bf Acknowledgement} This work was supported 
by U.S. DOE under DE-SC0020127, DE-FC02-04ER54698, and
by Subcontract S015879 with Princeton Plasma Physics Laboratory.
The help of R. Sweeney with the
DIII-D data and for discussions is acknowledged.

{\bf Data Availability} The data that support the findings of this study are available 
from the corresponding author upon reasonable request.


\begin{thebibliography}{9}
\bibitem{jet21}
H. Strauss and JET Contributors,
Effect of Resistive Wall on Thermal Quench in JET Disruptions,
Phys. Plasmas \textbf{28}, 032501 (2021) 
\bibitem{iter21} H. Strauss, Thermal quench in ITER disruptions,
 Phys. Plasmas \textbf{28} 072507 (2021) 

\bibitem{devries12}
P.C. de Vries, M.F. Johnson, B. Alper, P. Buratti,
T.C. Hender, H.R. Koslowski, V. Riccardo and JET-EFDA Contributors,
Survey of disruption causes at JET, Nucl. Fusion \textbf{51}   053018  (2011).
\bibitem{gerasimov2020}
S.N. Gerasimov,
 P. Abreu, G. Artaserse, M. Baruzzo, P. Buratti, I.S. Carvalho, I.H. Coffey,
E. De La Luna, T.C. Hender, R.B. Henriques, R. Felton, S. Jachmich, 
U. Kruezi, P.J. Lomas,
P. McCullen, M. Maslov, E. Matveeva, S. Moradi,
L. Piron1, F.G. Rimini, W. Schippers, C. Stuart, G. Szepesi, M. Tsalas,
D. Valcarcel, L.E. Zakharov and JET Contributors,
Overview of disruptions with JET-ILW,
Nucl. Fusion \textbf{60}  066028 (2020).
\bibitem{devries16}
P.C. de Vries, G. Pautasso, E. Nardon, P. Cahyna, S. Gerasimov,
J. Havlicek, T.C. Hender, G.T.A. Huijsmans, M. Lehnen, M. Maraschek, T. Markovič,
J.A. Snipes and the COMPASS Team, the ASDEX Upgrade Team and JET Contributors
Scaling of the MHD perturbation amplitude required to 
trigger a disruption and predictions for ITER,
Nucl. Fusion \textbf{56} 026007 (2016). 
\bibitem{schuller}
F.C. Schuller, Disruptions in tokamaks, Plasma Phys. Controlled Fusion \textbf{37}, A135 (1995).
\bibitem{pucella} G. Pucella, P. Buratti, E. Giovannozzi, E. Alessi,
F. Auriemma, D. Brunetti, D. R. Ferreira, M. Baruzzo,
D. Frigione, L. Garzotti, E. Joffrin, E. Lerche, P. J. Lomas, S. Nowak, L. Piron,
F. Rimini, C. Sozzi, D. Van Eester, and JET Contributors,
Tearing modes in plasma termination on JET:
the role of temperature hollowing and edge cooling,
 Nucl. Fusion \textbf{61} 046020 (2021)
\bibitem{sweeney} R. Sweeney,
W. Choi, M. Austin, M. Brookman, V. Izzo, M. Knolker, R.J. La Haye,
A. Leonard, E. Strait, F.A. Volpe and The DIII-D Team,
Relationship between locked modes and thermal quenches in DIII-D,
Nucl. Fusion \textbf{58}, 056022 (2018)

\bibitem{sweeney2017} R. Sweeney, W. Choi, R. J. La Haye, S. Mao, K. E. J.
Olofsson, F. A. Volpe, and the DIII-D Team,
Statistical analysis of m/n = 2/1 locked and quasi - stationary modes
with rotating precursors in DIII-D,
Nucl. Fusion 57 0160192 (2017)
\bibitem{lehnen}
M.Lehnen, K.Aleynikova, P.B.Aleynikov, D.J.Campbell, P.Drewelow, N.W.Eidietis,
Yu.Gasparyan, R.S.Granetz, Y.Gribov, N.Hartmann, E.M.Hollmann,  V.A.Izzo, S.Jachmich,
S.-H.Kim, M.Kočan, H.R.Koslowski, D.Kovalenko, U.Kruezi, A.Loarte, S.Maruyama,
G.F.Matthews, P.B.Parks, G.Pautasso, R.A.Pitts, C.Reux, V.Riccardo, R.Roccella,
J.A.Snipes, A.J.Thornton, P.C.de Vries, EFDA JET contributors,
Disruptions in ITER and strategies for their control and mitigation,
Journal of Nuclear Materials, \textbf{463}, 39 (2015)
\bibitem{finn95} John A. Finn,
 Resistive wall stabilization of kink and tearing modes
Phys. Plasmas 2, 198 (1995)
\bibitem{gimblett} C.G. Gimblett,
On free boundary instabilities induced by a resistive wall,
Nucl. Fusion  26, 617 (1986)
\bibitem{bondeson} A. Bondeson and M. Persson,
Stabilization by resistive walls and q-limit disruptions in tokamaks,
Nucl. Fusion 28, 1887 (1988)
\bibitem{betti} R. Betti,
Beta limits for the n = 1 mode in rotating - toroidal - resistive plasmas
surrounded by a resistive wall, Phys. Plasmas 5, 3615 (1998).

\bibitem{m3dc1} S. C. Jardin, N. Ferraro, J. Breslau, J. and Chen, Comput. Sci. \& Disc.,
 5, 014002 (2012)
\bibitem{ferraro}
N.M. Ferraro, S. C. Jardin, L. L. Lao, M. S. Shephard, and F. Zang,
Multi - region approach to free - boundary three - dimensional tokakmak equilibria
and resistive wall instabilities, Phys. Plasmas \textbf{23}, 056114 (2016). 
 \bibitem{m3d}
 W. Park, E.  Belova, G. Y.   Fu, 
X.  Tang, H. R.  Strauss, L. E.  Sugiyama,
Plasma Simulation Studies using Multilevel Physics Models,
  Phys. Plasmas \textbf{6} 1796 (1999).
\bibitem{pletzer} A. Pletzer and H. Strauss, Comput. Phys. Commun. 182, 2077 (2011).
\bibitem{frs73} H. P. Furth, P. H. Rutherford, and H. Selberg,
 Tearing mode in the cylindrical tokamak,
Physics of Fluids 16, 1054 (1973)
\end{thebibliography}
\end{document}